\documentclass[twoside,journal]{IEEEtran}

\usepackage{graphicx}
\usepackage{subfig}
\graphicspath{figures}
\usepackage{tabu}
\usepackage{multirow}
\usepackage{color}
\usepackage{tikz}
\usepackage{subfloat}
\usepackage{float}

\usepackage{amsmath}
\usepackage{amssymb}

\usepackage[space]{cite} 

\definecolor{corrAH}{rgb}{0,0,1}

\begin{document}

\title{Human Gaze Guided Attention for Surgical Activity Recognition}

\author{Abdishakour~Awale and
		Duygu~Sarikaya
	

}

\markboth{IEEE TRANSACTIONS ON MEDICAL IMAGING,~Vol.~X, No.~Y, MM~YYYY}{}

\maketitle

\begin{abstract}
Modeling and automatically recognizing surgical activities are fundamental steps
toward automation in surgery and play important roles in providing timely feedback to surgeons. Accurately recognizing surgical activities in video poses a challenging problem that requires an effective means of learning both spatial and temporal dynamics. Human gaze and visual saliency carry important information about visual attention and can be used to extract more relevant features that better reflect these spatial and temporal dynamics. In this study, we propose to use human gaze with a spatio-temporal attention mechanism for activity recognition in surgical videos. Our model consists of an I3D-based architecture, learns spatio-temporal features using 3D convolutions, as well as learns an attention map using human gaze as supervision. We evaluate our model on the \emph{Suturing} task of JIGSAWS which is a publicly available surgical video understanding dataset. To our knowledge, we are the first to use human gaze for surgical activity recognition. Our results and ablation studies support the contribution of using human gaze to guide attention by outperforming state-of-the-art models with an accuracy of 85.4\%.
 
\end{abstract}


\section{Introduction}  \label{Introduction}
Robot-assisted surgery allows surgeons to carry out a wide range of operations in a minimally invasive manner \cite{gao2020automatic}. With the help of robotic surgical systems, we have access to a substantial amount of video feed that we can use to model surgical activities. Automatically recognizing surgical activities is a fundamental step
toward automation in surgery and plays an important role in providing timely feedback and objective performance assessment to the surgeons in training. Surgical activities can be broken down into different hierarchical levels of activities such as the higher levels of activities; phases and steps, and the atomic level of activities called gestures. Surgical gestures are recurring common activity segments such as \emph{Positioning the needle, Pushing the needle through the tissue} within the context of surgical tasks \cite{gao2014jhu}. Accurately recognizing these activities in video poses a challenging problem that requires an effective means of learning both spatial and temporal dynamics. 

Frame-based image cues have been widely used for the recognition of surgical activities, however, these models can depend on irrelevant features that are specific to the environment of the task being performed rather than the surgical activities themselves. So, generalizability across different tasks and datasets remains a challenge \cite{sarikaya2020towards}.

Using attention mechanism \cite{vaswani2017attention} with deep learning models has shown a great performance boost in various tasks as they are able to focus on the more relevant features. These models learn attention distribution which is then used for feature re-weighting so that the features that are more relevant to the task are given higher importance. We see these advances in recent works that address human activity recognition. For example, Sudhakaran \emph{et al.} \cite{sudhakaran2019lsta} developed a Long Short Term Attention (LSTA) mechanism, which focuses on the more relevant spatial features for the problem of activity recognition in egocentric videos. Lu \emph{et al.} \cite{lu2019deep} proposed an attention-based two-stream deep neural network with motion and appearance streams that exploits the spatial and temporal dynamics of egocentric videos for activity recognition. 

On the other hand, human gaze patterns and visual saliency carry important information about the visual attention of humans. In other words, they provide a natural means of an attention mechanism and can be used to extract more relevant features that reflect these spatial and temporal dynamics in alignment with the human visual system. Deep learning models which utilize this information have proved to be effective in recognizing human activities, particularly in egocentric videos \cite{li2018eye}.  

In this paper, we propose to use the human gaze with a spatio-temporal attention mechanism for surgical activity recognition in videos. The motivation is that during object manipulation, a substantial amount of gaze fixation falls on the parts of objects which are relevant to the task that is being performed \cite{lu2019learning}. Therefore, focusing on these areas reduces misleading features extracted from the cluttered background and non-task-dependant objects that are not relevant to the activities being carried out.

Our model is an I3D-based architecture \cite{carreira2017quo} that learns attention maps using human gaze as supervision. It also learns spatio-temporal features from a sequence of consecutive video frames with 3D convolutions. First, we predict gaze points for each frame of the videos in the JIGSAWS dataset \cite{gao2014jhu}, which is a publicly accessible surgical video understanding datase, using a visual saliency prediction model \cite{huang2018predicting} trained on the GTEA Gaze \cite{fathi2012gteagaze} and GTEA Gaze+ \cite{li2015gteagazeplus} datasets. Next, we incorporate an attention module, similar to the module proposed by Lu \emph{et al.} \cite{lu2019learning}, which learns visual saliency with gaze information, into our I3D based architecture for surgical activity recognition. The spatio-temporal attention module consists of Inception blocks with 3D convolutions that learn spatio-temporal attention using human gaze information as supervision \cite{lu2019learning}. We evaluated our model on the \emph{Suturing} task of the JIGSAWS dataset. Our model outperforms state-of-the-art models with an accuracy of 85.4\% in this task. To the best of our knowledge, we are the first to propose using human gaze for surgical activity recognition. 

The contribution of our paper can be summarized as follows: 
\begin{itemize}
  \item To our knowledge, we are the first to use human gaze information for surgical activity recognition. 
  \item We incorporate a human gaze-guided spatio-temporal attention module in an I3D model for surgical activity recognition. Our approach outperforms state-of-the-art methods for the \emph{Suturing} task of the JIGSAWS dataset. 
  \item We provide an analysis of our model's results using different performance metrics and demonstrate how our proposed method using human gaze contributes to better surgical activity recognition.
\end{itemize}

The details of the paper are described as follows:  in the second section of the paper, related studies are presented, in the third section, details about the JIGSAWS, GTEA Gaze, and GTEA Gaze+ datasets are shown, in the fourth section, the proposed method is introduced, in the fifth section, experiments, implementation details, and results are discussed, in the sixth section, an analysis of the limitations and discussion on future works are illustrated and in the seventh section, we discuss the findings of our study.

\begin{figure*} [!t!b!h]
	\centering
	\includegraphics[width=\textwidth]{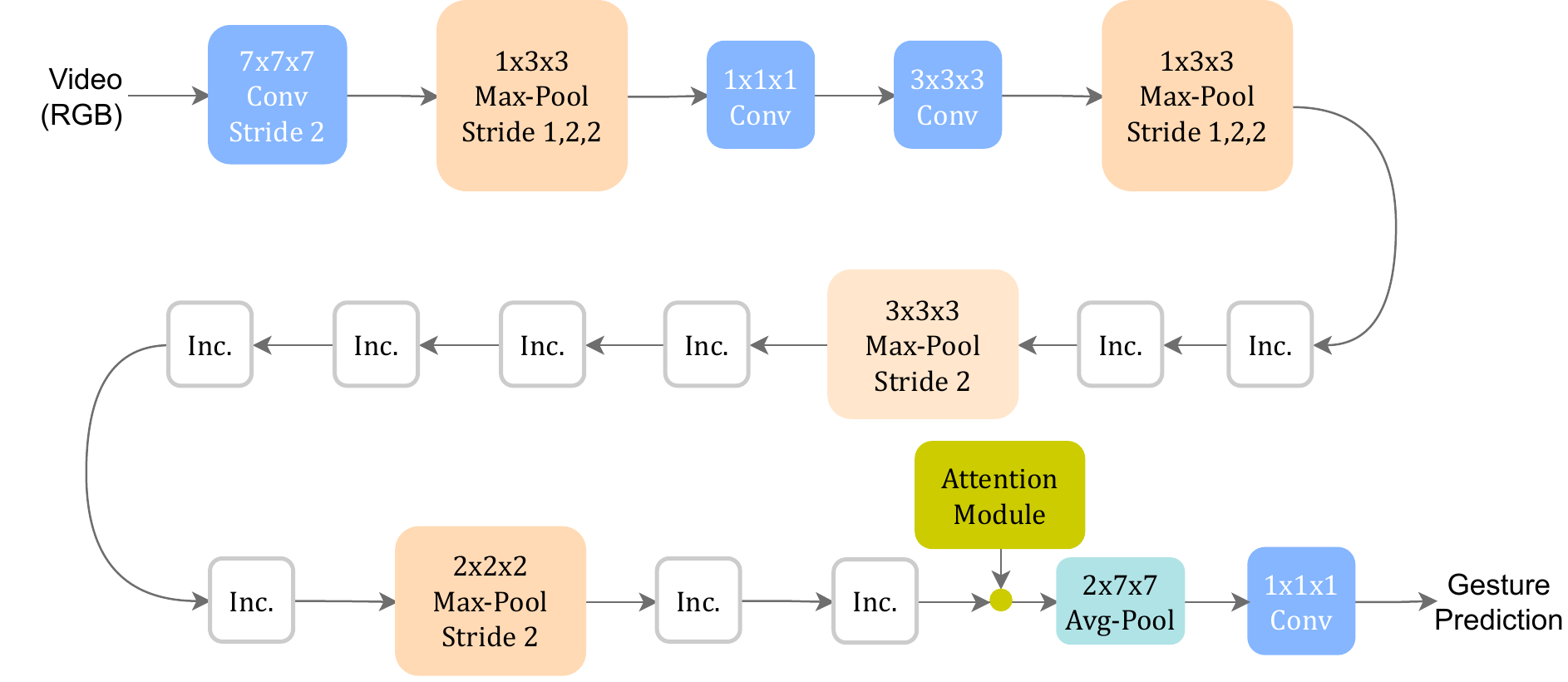}
	\caption{Our surgical activity recognition model is an I3D-based architecture and uses sequences of consecutive RGB video frames as input. A spatio-temporal attention module that uses human gaze as supervision is incorporated into this model to help the network selectively focus on salient parts of the images which are often related to the surgical activities being performed.}
	\label{fig:i3d}
\end{figure*}

\section{Related Work}  \label{RelatedWork}
\subsection{Surgical Activity Recognition in Video}

Recognizing activities in surgical videos is an active research area. The studies carried out in this domain range from phase recognition \cite{tran2017phase} to fine-grained gesture recognition \cite{Huaulme2021}. Earlier approaches were based on probabilistic models that used video frames or kinematics information as input. For example, a combined Markov/semi-Markov conditional random field (MsM-CRF) model has been utilized for joint segmentation and recognition of surgical gestures from kinematic and video data \cite{tao2013surgical}. 

More recently, Deep Neural Networks (DNNs) have been widely adopted for surgical activity recognition. Zia \emph{et al.} \cite{zia2018surgical} proposed to use an RP-Net (Reverse Projection Network) which is based on a modified version of the Inception-V3 architecture to recognize activities in surgical videos obtained from robot-assisted radical prostatectomies. The proposed system consists of two components: System data and video data models. They utilized multiple RNNs to learn temporal dynamics from instrument motion for activity recognition, while they used the Inception-V3 network for extracting visual features from robot-assisted radical prostatectomies video data. Amsterdam \emph{et al.} Zisimopoulos \emph{et al.} \cite{zisimopoulos2018deepphase} proposed to use a CNN along with an RNN to recognize surgical phases in cataract surgeries. They trained a Resnet-152 model to recognize surgical tools used during cataract surgeries. The Resnet-152 learns to classify 21 distinct tool classes. The output of the Resnet-152 is then processed by an RNN network for phase recognition. The RNN learns temporal features and tool presence in video frames to recognize which surgical phases are being carried out.

Czempiel et al. \cite{czempiel2020tecno} proposed a multi-stage temporal convolution network that consists of two modules that jointly learn spatial and temporal features from surgical videos for surgical phase recognition. The proposed method has two major submodules: a feature extraction module, which is based on ResNet-50 architecture and extracts and learns both high and low-level spatial features from surgical video frames, and a temporal module, which is based on Temporal Convolution Network (TCN) that extracts temporal dynamics from stacks of video frames. The extracted visual features by the visual extractor are fed as input to the temporal module of the multi-stage network. The proposed approach is tested on the two different publicly available laparoscopic cholecystectomy procedure datasets.  

Funke \emph{et al.} \cite{funke2019using} learned spatiotemporal features from stacks of video frames with 3D CNNs to recognize activities in surgical videos. Instead of relying only on spatial features extracted with 2D Convolutions, they proposed applying 3D convolutions and 3D kernels to consecutive video frames to extract both spatial and temporal features. 

Inspired by the wide adoption of 3D CNNs and their outstanding abilities in extracting both visual and temporal data from volumes of consecutive video frames, Ding \emph{et al.} \cite{ding2020surgical} proposed to use a two-stream network that jointly learns spatial and temporal features from surgical videos for surgical phase recognition. The proposed model, which is named a Two-Stream Mixed Convolutional Network (TsMCNet), learns spatiotemporal features from surgical video frames and recognizes which surgical phase is taking place. The TsMCNetmodel consists of three major components: Shared CNNs for learning visual representations, Visual Branch for extracting spatial features and Temporal Branch for learning temporal features. The proposed method is evaluated on MICCAI 2016 Workflow Challenge dataset and achieved promising results. 

Huynhnguyen \emph{et al.} \cite{huynhnguyen2021toward} proposed a hybrid deep learning model that learns and classifies surgical gestures in a two-stage process. In the first stage, a 3D CNN model is utilized to determine whether a 10-consecutive-frame input belongs to the same surgical task or is a transition between two distinct surgical gestures. In the second stage of the experiment, a hybrid network that consists of LSTM and CNN models is used to classify surgical video segments into ten different gestures. The latter network is utilized for learning spatiotemporal features from consecutive video frames. They evaluated their approach on the Suturing task of the JIGSAWS dataset by following the Leave-One-Supertrial-Out (LOSO) experimental setup.

Following the superior performance of the 3D CNNs in surgical activity recognition as shown by Funke \emph{et al.} \cite{funke2019using}, we use an I3D-based architecture with our human gaze-guided attention mechanism. The idea of Inflated 3D ConvNet (I3D) was first proposed in \cite{carreira2017quo} and is a conflated 2D ConvNet. Filters and pooling kernels of very deep image classification ConvNets are expanded into 3D to learn spatio-temporal features from videos. The I3D learns spatial and temporal dynamics from volumes of video frames, and has widely been utilized for action recognition tasks thanks to its superior performance in activity recognition.

\subsection{Attention Mechanisms for Surgical Activity Recognition }
Zhang \emph{et al.} \cite{zhang2020symmetric} proposed a symmetric dilation convolution network that is incorporated in a self-attention module for surgical activity recognition. The proposed model learns spatial and temporal dependencies directly from surgical video input. They first extract spatial features with a Temporal Convolutional Network (TCN) and use the extracted features as input for the symmetric dilation network. Temporal features are further learned with a sequence of 1D dilated convolutions. Furthermore, an attention mechanism is then utilized to capture latent representations from video frames. Contrary to learning features from consequent video frames forming segments with 3D CNNs, they consider taking the entirety of the surgical video as input to the network. They demonstrated the effectiveness of their approach on the Suturing task of JIGSAWS dataset.

\subsection{Human Gaze Based Attention Models for Human Activity Recognition}
Attention-based deep learning models have recently been used for human action recognition. Attention mechanisms are incorporated into deep learning models to guide them to focus on informative and salient regions in visual data and facilitate the task of activity recognition. Sudhakaran \emph{et al.} \cite{sudhakaran2018attention} learn highly specialized attention maps for each video frame with an end-to-end trainable deep neural model for egocentric action recognition. The proposed model is a CNN-RNN architecture and uses the spatial attention mechanism to enable the network to selectively focus on salient parts of the visual input. Du \emph{et al.} \cite{du2017recurrent} proposed a novel recurrent spatial-temporal attention network (RSTAN) for action recognition in videos, which adaptively learns spatio-temporal features from video frames for better action recognition. The proposed attention module is incorporated into an LSTM model to identify spatial-temporal features which are strongly relevant to the activities in videos. Lu \emph{et al.} \cite{lu2019learning} introduced a spatio-temporal attention module that predicts attention maps with human gaze information as supervision for recognizing egocentric activities. The attention module is augmented in a two-stream I3D network where each stream has its own attention map predicting module to help the network focus on the relevant image regions for action recognition. Huang \emph{et al.} \cite{huang2020mutual} proposed a Mutual Context Network (MCN), which is a deep learning-based framework, to jointly model gaze prediction and action recognition in egocentric videos. Min \emph{et al.} \cite{min2021integrating} proposed a probabilistic approach to integrating human gaze into spatio-temporal attention for egocentric activity recognition. They present gaze locations as structured discrete latent variables to model their uncertainties. The modeled gaze distribution is then used to guide the network to the relevant image regions for better action recognition.

In this paper, we propose to utilize human gaze with a spatio-temporal attention module that learns visual attention from feature volumes for surgical activity recognition. Our approach is based on the popular I3D architecture and learns attention maps with a spatio-temporal attention module using human gaze as supervision. First, we predict gaze points for each video frame in the \emph{Suturing} task of the JIGSAWS dataset using a visual saliency prediction model \cite{huang2018predicting} trained on the GTEA Gaze \cite{fathi2012gteagaze} and GTEA Gaze+ \cite{li2015gteagazeplus} datasets. The generated gaze points are used as ground-truth gaze information for the spatio-temporal attention module during model training. We incorporate the attention module, proposed by Lu \emph{et al.} \cite{lu2019learning}, which learns visual saliency with gaze information, into our I3D model for surgical activity recognition. To our knowledge, we are the first to use human gaze with an attention module for surgical activity recognition.

\section{Dataset} \label{Dataset}
\subsection{JIGSAWS Dataset}
The JHU-ISI Gesture and Skill Assessment Working Set (JIGSAWS) \cite{gao2014jhu} is a publicly available surgical activity dataset. This dataset contains recordings of surgical videos obtained with the endoscopic camera of the da Vinci Surgical System. In these videos, $8$ different surgeons with different surgical skill levels perform surgical tasks. There are three main tasks in this dataset: \emph{Suturing, Needle Passing} and \emph{Knot Tying.} The dataset also includes gesture labels such as \emph{positioning the needle, pushing the needle through the tissue.} As can be seen from table \ref{tablo2}, there is a total of $10$ different surgical gestures in the \emph{Suturing} task of the JIGSAWS dataset. We follow the standardized leave-one-user-out (LOUO) experimentation setup during our experiments. 
\begin{table}[hbt!]
  \centering
   \caption{Gesture Vocabulary}
   
  \label{tablo2}
  \begin{tabular}{| m{6em} | m{22em}|}
    \hline
    Index & Gesture Description \\
    \hline
    G1 & Reaching for needle with right hand \\
    \hline
    G2 & Positioning needle \\
    \hline
    G3 & Pushing needle through tissue \\
    \hline
    G4 & Transferring needle from left to right \\
    \hline
    G5 & Moving to center with needle in grip \\
    \hline
    G6 & Pulling suture with left hand \\
    \hline
    G8 & Orienting needle \\
    \hline
    G9 & Using right hand to help tighten suture \\
    \hline
    G10 & Loosening more suture \\
    \hline
    G11 & Dropping suture at end and moving to end points \\
    \hline
    
  \end{tabular}
\end{table}

\subsection{Gaze Dataset}
We use human gaze supervision in order to guide the training of the spatio-temporal attention module for effective surgical activity recognition. Since the JIGSAWS dataset doesn't come with human gaze data, we utilize a visual saliency gaze prediction model which was pretrained on the GTEA Gaze \cite{fathi2012gteagaze} and GTEA Gaze+ \cite{li2015gteagazeplus} datasets to predict gaze locations for our frames. The predicted gaze points focus on the salient parts such as the surgical tools and the suture which are often relevant to the surgical activities that take place. Even though the GTEA Gaze \cite{fathi2012gteagaze} and GTEA Gaze+ \cite{li2015gteagazeplus} datasets were designed for egocentric activities in first-person videos recorded with a wearable camera, we argue that they are similar to surgical videos recorded with endoscopic cameras as the field of operation is captured with a moving camera focused on this field. The results also support this argument. As can be seen in figure \ref{fig:heatmaps}, the predicted gaze locations accurately reflect the activities that are being carried out and focus on relevant parts such as the surgical tools and the suture that is key to recognizing which surgical gesture is taking place. The gaze locations are represented as (x, y) coordinates on each frame and are used during model training. The gaze locations are used as supervision for the spatio-temporal attention module to learn attention and generate attention maps. 

\begin{figure}[!h]
	\centering
	\includegraphics[width = \columnwidth]{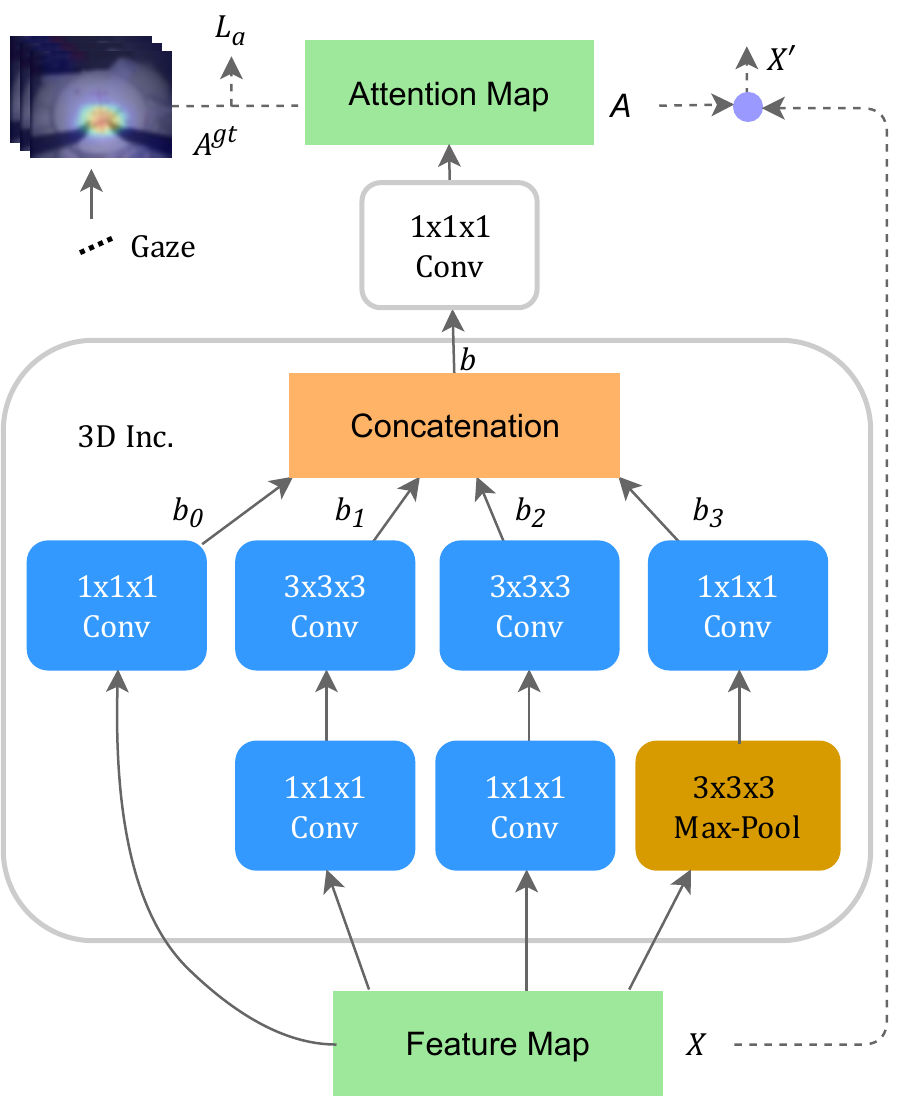}
	\caption{ The spatio-temporal attention module uses a 3D Inception (3D Inc.) and a 3D convolution layer to predict attention maps.The spatio-temporal attention module learns attention maps to help the network selectively focus on the relevant parts of the video frames to recognize surgical activities. Human gaze information is used as supervision for learning attention maps.}
	\label{fig:stam}
\end{figure}

\section{Methods} \label{Methods}

\subsection{Spatio-temporal Attention Module }
Our model is an I3D-based architecture that learns spatio-temporal features with 3D convolutions. In addition, we adopt the spatio-temporal attention module proposed by Lu \emph{et al.} \cite{lu2019learning}. The spatio-temporal attention module learns attention maps to help the network selectively focus on the relevant parts of the video frames to recognize surgical activities. Human gaze information is used as supervision for learning attention maps.

\begin{figure}[!h]
	\centering
	\includegraphics[width = \columnwidth]{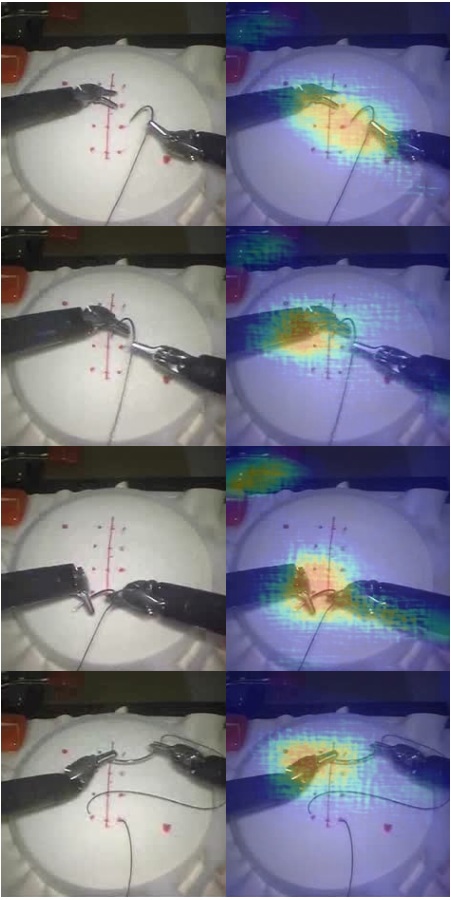}
	\caption{ Heat maps based on estimated gaze points for a few sample video frames from the JIGSAWS dataset are shown. The predicted gaze locations are consistent with human eye gaze fixation and accurately represent the surgical activity that are being carried out. }
	\label{fig:heatmaps}
\end{figure}

To use the spatio-temporal attention module for surgical activity recognition, we first predict the gaze points for each frame of the videos in the JIGSAWS dataset with a visual salience prediction model \cite{huang2018predicting} trained on the GTEA Gaze \cite{fathi2012gteagaze} and GTEA Gaze+ \cite{li2015gteagazeplus} datasets. Although GTEA Gaze and GTEA Gaze+ are designed for action recognition in egocentric videos, we argue that they are similar to surgical activities recorded with an endoscopic camera. For this reason, we observe that the surgical activity videos are compatible with the human gaze predicted with model. The figure \ref{fig:heatmaps} shows heat maps based on estimated gaze points for a few sample video frames from the JIGSAWS dataset.

The spatio-temporal attention module shown in figure \ref{fig:stam} consists of 3D Inception module and 3D convolutions \cite{lu2019learning}. It takes feature maps \emph{X} as input and outputs attention maps \emph{A}.

\begin{equation}
  X
    \in
    R^{C\times T \times H\times W}
\end{equation}
\begin{equation}
  A
    \in
    R^{T \times H\times W}
\end{equation}

The \emph{C} in the feature maps denotes the input channels. The attention module processes feature maps and produces attention maps as follows: 

\begin{equation}
  \begin{aligned}
   & b0 = conv1\_0(X)\\
   & b1 = conv3\_1(conv1\_1(X))\\
   & b2 = conv3\_2(conv1\_2(X))\\
   & b3 = conv1\_3(max pool(X))\\
   & b = concat([b0; b1; b2; b3])\\
   & A = f(conv1\_a(b)),
  \end{aligned}
\end{equation}

\emph{conv3\_i} denotes 3D convolution with 3 × 3 × 3 kernel, \emph{conv1\_i} denotes 3D convolution with 1 × 1 × 1 kernel, and \emph{f} represents a linear function that scales the input to $[0, 1]$ \cite{lu2019learning}. There are four branches in the 3D Inception module $(b0, b1, b2, b3)$ which produce intermediate feature maps.  The intermediate feature maps are then concatenated and processed by the last convolutional layer \emph{(conv1\_a)} and the scale function \emph{f} to produce the spatiotemporal attention map A. The predicted feature map A is a 3D feature volume with resolution \emph{T, H, W}, which consists of a spatial attention for each time stamp. With 3D convolutions, we simultaneously learn both spatial and temporal features of sequences of consecutive RGB video frames to predict attention maps.

To guide the training of the 3D attention module, we use predicted gaze points generated with a gaze prediction model \cite{huang2018predicting}. The predicted attention map A is combined with the input feature map X to produce a more relevant feature map \(X'\). The generated attention maps illustrate the regions where our model is actually focusing on and can be seen that these regions are relevant to the activities taking place. Higher weights are given to the \(X'\) feature map and is used as an input to the next layers in the network for surgical activity recognition.

\subsection{Network Architecture}
Our activity recognition model is based on the popular I3D architecture \cite{carreira2017quo} which is widely used for action recognition tasks. The I3D model extends Inception-V1 to be a 3D CNN. In this study, our I3D-based model uses sequence of RGB video frames as input and learns spatio-temporal features with 3D convolutions \cite{carreira2017quo}. The I3D is made by converting 2D CNNs into 3D ConvNets \cite{carreira2017quo}. The filters
and pooling kernels of a deep 2D architecture are inflated to create a 3D CNN with $N \times N \times N$ dimensions, giving them the ability to extract both spatial and temporal dynamics from stacks of video frames. Carreira \emph{et al.} \cite{carreira2017quo} bootstrap 3D from a pretrained 2D CNN model and they used the parameters of these pretrained 2D CNN models, which were trained on very large datasets such as ImageNet dataset \cite{imagenet}, for their 3D CNN implementation. To put this into perspective, they repeat the weights of 2D filters $N$ times along the temporal dimension and then they rescale them back by dividing $N$ times \cite{carreira2017quo}. To ensure the accuracy of this operation, filters of average and max pooling are kept the same for 3D CNN as to those of the 2D CNN. 
Another modification that is essential to the I3D model is the receptive field in the convolutional and pooling layers. The receptive field in deep learning models is the part of the input image that is visible to one filter at a time during the convolution operation, and it grows bigger as we stack more hidden layers. In 2D CNNs, the convolution filters and pooling kernels focus on the height and width of the 2D input. And as a result, their filters and pooling kernels are symmetrical. However, the addition of the third dimension, which is the time in this case, makes the convolution filters and pooling kernels of the I3D model asymmetric. The asymmetric structure of 3D CNNs facilitates the capture of spatiotemporal features from stacks of video frames. This network was designed in such a way that it grows wider instead of deeper to allow the extraction of spatiotemporal information at various scales and then it aggregates the results. The I3D model was pretrained on the Kinetics dataset \cite{kinetics}, which is a large dataset that contains labeled videos of human actions. In our study, we incorporated a spatiotemporal attention module into our I3D model. The attention module learns and predicts attention maps to help the network focus on highly relevant spatial regions for surgical activity recognition.

As noted in figure \ref{fig:i3d}, a spatio-temporal attention module is included in this model. The spatio-temporal attention module predicts human gaze-guided attention maps, then the predicted attention map is combined with the input feature map to produce a more relevant feature map with weighted average pooling \cite{lu2019learning}.



\section{Experiments} \label{Experiments}
\subsection{Implementation Details}
We train our I3D model with segments taken from the trial videos as sliding windows to predict the corresponding ground-truth surgical gesture labels. Input to the network is a stack of 16 consecutive frames with a resolution of 224x224 pixels. We also perform data augmentation by randomly flipping video frames horizontally. The gaze location is adjusted according to the data augmentation performed on the video frames. Weights of an I3D model pretrained on the Imagenet and Kinetics dataset \cite{carreira2017quo} is used as an initialization during model training. We train our network with a batch size of 12 with Stochastic Gradient Descent (SGD) optimization algorithm with a 0.9 momentum, a 0.0000007 weight decay, and an initial learning rate of 0.1. The learning rate decreases by a factor of 0.1 after 1k iterations and the model is trained for 10k iterations. We also set the dropout rate to 0.5. We implemented our model with the Pytorch framework.

\subsection{Evaluation}
We evaluated our model on all the video snippets in the 39 videos found in the \emph{Suturing} task of the JIGSAWS dataset and we used the Leave-One-User-Out (LOUO) experimental setup for cross-validation. In the LOUO, all of the trials that belong to a one user are left out for testing while the remaining are used for training the model. Accordingly, we report the average accuracy of all the experiments we performed for eight different surgeons in Table \ref{results_table}. we also report Average \emph{F1} score, where we calculate the harmonic mean of precision and recall.

We compared our results with the state-of-the-art 3D CNN methods and an earlier baseline method using 2D ResNet18 on the JIGSAWS benchmark. Please note that the "window" in the benchmarks refers to a look ahead at future frames in order to achieve more accurate activity recognition. Even though we do not look ahead at future frames, we achieve superior results compared to the benchmark studies, and we perform significantly better compared to the 3D CNN model suggested by \cite{funke2019using} without a look-ahead window. The results also show how effective our human gaze-guided attention mechanism works compared to the baseline I3D model without attention. We achieved an average accuracy of 85.4\% on the \emph{Suturing} task, outperforming state-of-the-art methods. Our approach suggests that using human gaze with spatio-temporal attention greatly improves model performance by guiding the network to identify the salient regions, hence helping the network predict surgical activities more accurately. The Edit score is also reported in \ref{results_table}, which demonstrates the quality of predicted video snippets.

\begin{table*}[!t!b!h]
  \centering
   \caption{A comparison of our model results with other methods on the \emph{Suturing} task is shown in this table. For cross-validation, we used the Leave-One-User-Out (LOUO) experimental setup. Accordingly, we reported the average accuracy of all experiments we performed for eight different surgeons. Evaluations are at 5 fps. Measures are given in \%. Even though we do not look ahead at future frames, we achieve superior results compared to the benchmark studies. The results also show how effective our human gaze-guided attention mechanism works compared to the baseline I3D model. }
  \label{results_table}
  \begin{tabular}{| m{28em}| m{8em}|  m{8em}|  m{8em} |}
    \hline
    Method & Average Accuracy & Average \emph{F1} & Edit Score \\
    \hline
    2D ResNet-18 \cite{funke2019using} & 79.9 & 73.3 & 41.4 \\
    \hline
    3D CNN \cite{funke2019using} & 79.9  & 73.7 & 64.0\\
    \hline
    3D CNN + window \cite{funke2019using} & 84.0 & 78.4 & 80.7\\
    \hline
    I3D model without attention & 77.6 & 70.1 & 72.8\\
    \hline
    Ours (I3D with human gaze guided attention) & \textbf{85.4 $\pm$  5.3} & \textbf{80.6} & \textbf{81.2}\\
    \hline
  \end{tabular}
\end{table*}



\subsection{Edit Distance} \label{Qualitative}
\begin{figure*}
    \centering
   \includegraphics[width=0.95\textwidth]{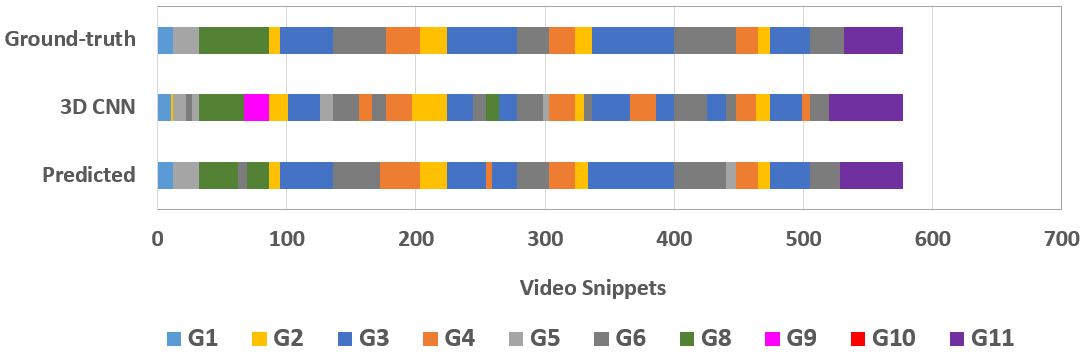}
  \caption{Edit distance analysis of the predicted results on the \emph{Suturing\_B002} video of Suturing task of the JIGSAWS dataset. Ground-truth labels are displayed above, the predictions by the benchmark 3D CNN model are shown in the middle, and the gestures predicted by our model are displayed at the bottom of the figure. }
  \label{qualitative}
\end{figure*}

\begin{figure*}
\centering
   \includegraphics[width=0.95\textwidth]{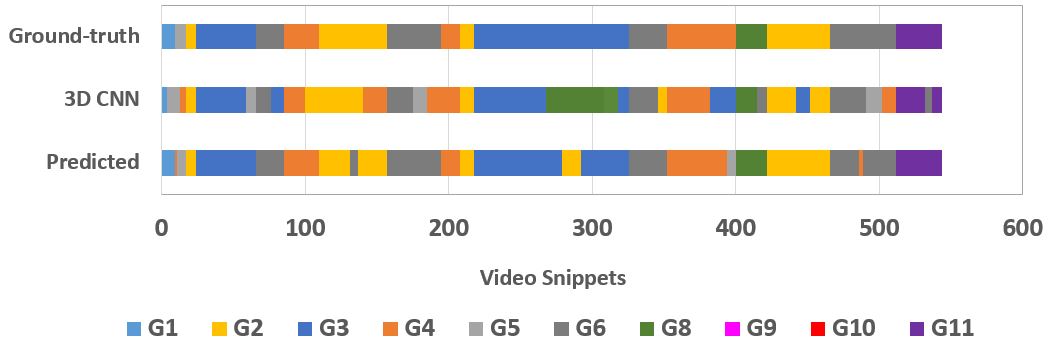}
  \caption{ Edit distance analysis of the predicted results on the \emph{Suturing\_B004} video of Suturing task of the JIGSAWS dataset. Ground-truth labels are displayed above, the predictions by the benchmark 3D CNN model are shown in the middle, and the gestures predicted by our model are displayed at the bottom of the figure.}
  \label{qualitative2}
\end{figure*}

\begin{figure*}
\centering
   \includegraphics[width=0.95\textwidth]{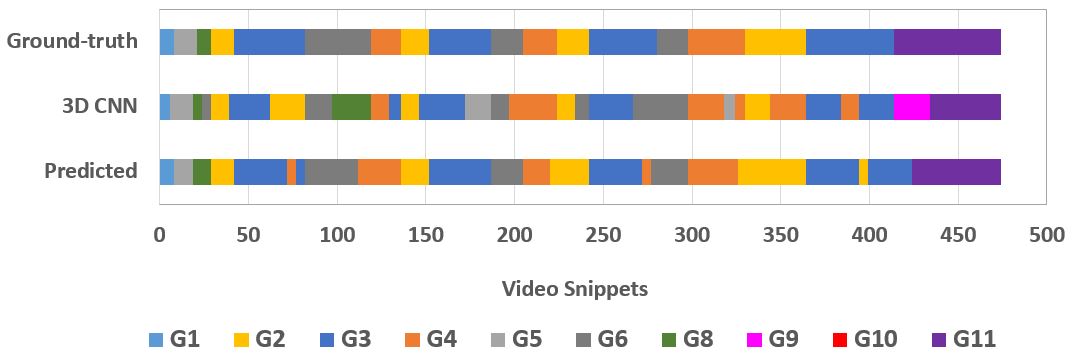}
  \caption{Edit distance analysis of the predicted results on the \emph{Suturing\_F001} video of Suturing task of the JIGSAWS dataset. Ground-truth labels are displayed above, the predictions by the benchmark 3D CNN model are shown in the middle, and the gestures predicted by our model are displayed at the bottom of the figure.
  }
  \label{qualitative3}
\end{figure*}

\begin{figure*}
\centering
   \includegraphics[width=0.95\textwidth]{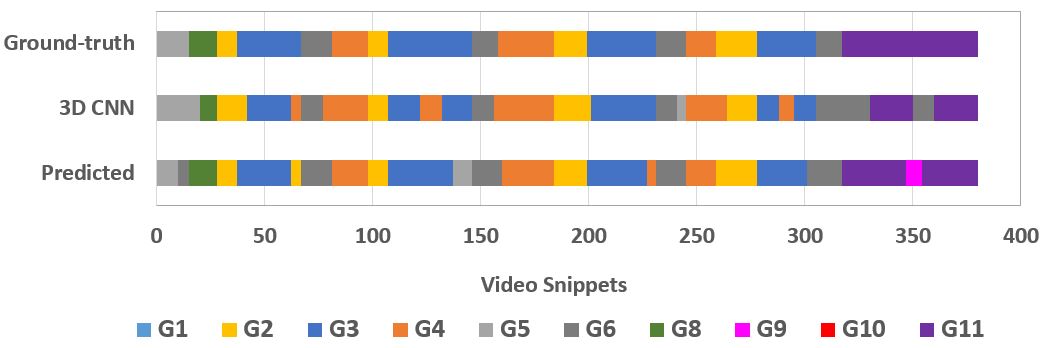}
  \caption{Edit distance analysis of the predicted results on the \emph{Suturing\_F003} video of Suturing task of the JIGSAWS dataset. Ground-truth labels are displayed above, the predictions by the benchmark 3D CNN model are shown in the middle, and the gestures predicted by our model are displayed at the bottom of the figure.}
  \label{qualitative4}
\end{figure*}

To analyze the effectiveness of our approach and how using human gaze with spatio-temporal attention module contributes to improved surgical activity recognition, we present the edit distance of our prediction trials which employs the Levenshtein distance to assess the quality of predicted video snippets. We demonstrate the superiority of our approach by comparing it with other state-of-the-art 3D CNN methods. As can be seen from the qualitative results in figures \ref{qualitative}, \ref{qualitative2}, \ref{qualitative3}, and \ref{qualitative4} our model demonstrates far better performance than the 3D CNN proposed in \cite{funke2019using}. While the benchmark 3D CNN model's predictions are cluttered, our method does a good job with the predictions; they are close to the ground truth gesture labels, the transitions between gestures are smoother and more in line with the surgical task workflow.



\section{Limitations and Future Work} \label{future}
A limitation of our study is that sometimes gaze locations could be misleading and uninformative as we used a human gaze model trained on GTEA Gaze \cite{fathi2012gteagaze} and GTEA Gaze+ \cite{li2015gteagazeplus} datasets. Although we argue that there are similarities to data captured with egocentric cameras, the ground-truth gaze points might not be directly correlated with the actual visual attention and as a result, could mislead the model. In order to make our model robust to this problem, a comprehensive study with recorded gaze information of the surgeons should be conducted. 


\section{Discussion and Conclusion} \label{Results}
Human gaze patterns and visual saliency carry important information about visual attention. Although state-of-the-art methods accepted in the literature learn spatio-temporal features for surgical activity recognition, none of them use visual saliency and human gaze.  To our knowledge, we are the first to use human gaze for surgical activity recognition. In our study, we proposed to use a human gaze-guided attention mechanism for recognizing surgical activities. To achieve this, we used a model with an I3D-based architecture, which uses a spatio-temporal attention mechanism that generates an attention map focusing on the salient parts using human gaze information as supervision. To implement the human gaze-guided spatio-temporal attention module for surgical activity recognition, we first predicted gaze points for each frame of the videos in the JIGSAWS dataset using a visual saliency prediction model \cite{huang2018predicting} trained on the GTEA Gaze \cite{fathi2012gteagaze} and GTEA Gaze+ \cite{li2015gteagazeplus} datasets. As can be seen in Figure \ref{fig:heatmaps}, the generated attention maps based on the estimated gaze points for a few sample video frames taken from the JIGSAWS dataset are consistent with the human gaze. Although GTEA Gaze \cite{fathi2012gteagaze} and GTEA Gaze+ \cite{li2015gteagazeplus} datasets were prepared for action recognition in egocentric videos, we observe that they are similar to surgical activities recorded with an endoscopic camera. We evaluated our model on the Suturing task of the JIGSAWS dataset and demonstrate an average accuracy of 85.4\% which suggests that learning visual attention with human gaze greatly improves model performance in surgical activity recognition. We also present an edit distance analysis of predicted video snippets to evaluate the quality of model predictions and we demonstrate the effectiveness of our method compared to the state-of-the-art approaches.




\nocite{*}
\bibliographystyle{IEEEtran}
\bibliography{humangaze.bib}

%

\end{document}